\PassOptionsToPackage{pdfpagelabels=false}{hyperref} 

\documentclass[fleqn,usenatbib]{mnras}

\usepackage{newtxtext,newtxmath}

\usepackage[T1]{fontenc}
\usepackage{ae,aecompl}

\usepackage[pdftex]{graphicx}
\usepackage{amsmath}	
\usepackage{threeparttable}
\usepackage{comment}
\usepackage{xspace}
\usepackage{hyperref}
\usepackage{url}
\usepackage{afterpage,array,rotating}
\usepackage{arydshln}
\usepackage{nccmath}
\usepackage{multirow}

\makeatletter
\newcommand{\figcaption}[1]{\def\@captype{figure}\caption{#1}}
\newcommand{\tblcaption}[1]{\def\@captype{table}\caption{#1}}
\makeatother

\newcolumntype{Y}{>{\centering\arraybackslash}p{4.5em}}
\newcommand{\nh}{\ensuremath{N_\mathrm{H}}\xspace}

\newcommand{\xmm}{\textit{XMM-Newton}\xspace}
\newcommand{\su}{\textit{Suzaku}\xspace}
\newcommand{\nus}{\textit{NuSTAR}\xspace}
\newcommand{\xmmnus}{\textit{XMM}-\textit{NuSTAR}\xspace}
\newcommand{\ngc}{NGC~5548\xspace}
\newcommand{\phabs}{{\tt phabs}\xspace}
\newcommand{\pl}{{\tt cutoffPL}\xspace}
\newcommand{\mekal}{{\tt mekal}\xspace}
\newcommand{\pexmon}{{\tt pexmon}\xspace}
\newcommand{\diskbb}{{\tt diskbb}\xspace}
\newcommand{\mtable}{{\tt mtable} \xspace}

\title[X-ray spectral variation of \ngc]{Simple interpretation of the seemingly complicated X-ray spectral variation of \ngc}

\author[Midooka T. et al.]{
Takuya Midooka$^{1,2}$,\thanks{E-mail: midooka@ac.jaxa.jp}
Ken Ebisawa$^{1,2}$,
Misaki Mizumoto$^{3,4}$,
Yasuharu Sugawara$^{1}$
\\
$^{1}$Institute of Space and Astronautical Science (ISAS), Japan Aerospace Exploration Agency (JAXA), 3-1-1 Yoshinodai, Chuo-ku, Sagamihara, Kanagawa 252-5210, Japan\\
$^{2}$Department of Astronomy, Graduate School of Science, The University of Tokyo, 7-3-1 Hongo, Bunkyo-ku, Tokyo 113-0033, Japan \\
$^{3}$Hakubi Center, Kyoto University, Yoshida-honmachi, Sakyo-ku, Kyoto, 606-8501, Japan\\
$^{4}$Department of Astronomy, Graduate School of Science, Kyoto University, Kitashirakawa-Oiwakecho, Sakyo-ku, Kyoto, 606-8502, Japan 
}
\date{Accepted XXX. Received YYY; in original form ZZZ}

\pubyear{2022}

\begin{document}
\label{firstpage}
\pagerange{\pageref{firstpage}--\pageref{lastpage}}
\maketitle

\begin{abstract}
\ngc is a very well-studied Seyfert 1 galaxy in broad wavelengths. Previous multi-wavelength observation campaigns have indicated that its multiple absorbers are highly  variable and complex. A previous study applied a two-zone partial covering  model with different covering fractions to explain the complex X-ray spectral variation and reported a correlation between one of the covering fractions and the photon index of the power-law continuum. However, it is not straightforward to physically understand such a correlation.
In this paper, we propose a model to avoid this unphysical situation; the central X-ray emission region is partially covered by \textit{clumpy absorbers composed of double layers}.
These ``double partial coverings'' have precisely the same covering fraction.
Based on our model, we have conducted an extensive spectral study using the data taken by \xmm, \su, and \nus in the range of 0.3--78~keV for 16 years. Consequently, we have found that the X-ray spectral variations are mainly explained by independent changes of the following three components; (1) the soft excess spectral component below $\sim$1~keV, (2) the cut-off power-law normalization, and (3) the partial covering fraction of the clumpy absorbers. In particular, spectral variations above $\sim$1~keV are mostly explained only by the changes of the partial covering fraction and the power-law normalization. In contrast, 
the photon index and all the other spectral parameters are not significantly variable.
\end{abstract}




\section{Introduction}

\begin{table*}
	\centering
	\caption{Observation logs with \xmm (pn), \su and \nus}
	\label{tab:obs_table}
	\begin{threeparttable}
	\begin{tabular}{ccccc}
		\hline
		Obs. Name & Satellite & Obs. ID & Start--End date [UT] & Exposure time (ks)\tnote{a} \\
		\hline
		*X1\tnote{b} & \xmm & 0109960101 & 2000 Dec. 24--25 & 22.9\\
		*X2 & \xmm & 0089960301 & 2001 Jul. 09--10 & 83.0\\
		*X3 & \xmm & 0089960401 & 2001 Jul. 12--12 & 28.1\\
		\hline
		S1 & \su& 702042010 & 2007 Jun. 18--19 & 31.1\\
		S2 & \su& 702042020 & 2007 Jun. 24--25 & 35.9\\
		S3 & \su& 702042040 & 2007 Jul. 08--08& 30.7\\
		S4 & \su& 702042050 & 2007 Jul. 15--15 & 30.0\\
		S5 & \su& 702042060 & 2007 Jul. 22--22 & 28.9\\
		S6 & \su& 702042070 & 2007 Jul. 29--29 & 31.8\\
		S7 & \su& 702042080 & 2007 Aug. 05--05 & 38.8\\
		\hline
		*X4 & \xmm & 0720110301 & 2013 Jun. 22--22 & 50.5\\
		*X5 & \xmm & 0720110401 & 2013 Jun. 30--30 & 55.1\\
		*X6 & \xmm & 0720110501 & 2013 Jul. 07--08 & 55.5\\
		*X7 & \xmm & 0720110601 & 2013 Jul. 11--12 & 55.4\\
		*X7N\tnote{c} & *\nus & 60002044002/3 & 2013 Jul. 11--12 & 55.2\\
		*X8 & \xmm & 0720110701 & 2013 Jul. 15--16 & 55.5\\
		*X9 & \xmm & 0720110801 & 2013 Jul. 19--20 & 56.5\\
		*X10 & \xmm & 0720110901 & 2013 Jul. 21--22 & 55.5\\
		*X11 & \xmm & 0720111001 & 2013 Jul. 23--24 & 55.5\\
		*X11N\tnote{c} & *\nus & 60002044005 & 2013 Jul. 23--24 & 53.1\\
		*X12 & \xmm & 0720111101 & 2013 Jul. 25--26 & 50.7\\
		*X13 & \xmm & 0720111201 & 2013 Jul. 27--28 & 55.5\\
		*X14 & \xmm & 0720111301 & 2013 Jul. 29--30 & 50.4\\
		*X15 & \xmm & 0720111401 & 2013 Jul. 31--Aug. 01 & 55.5\\
		*X16 & \xmm & 0720111501 & 2013 Dec. 20--21 & 55.3\\
		*X16N\tnote{c} & *\nus & 60002044008 & 2013 Dec. 20--21 & 53.8\\		
		*X17 & \xmm & 0720111601 & 2014 Feb. 04--05 & 55.4\\
		X18 & \xmm & 0771000101 & 2016 Jan. 14--14 & 35.2\\
		X19 & \xmm & 0771000201 & 2016 Jan. 16--16 & 32.5\\
		\hline
	\end{tabular}
	\begin{tablenotes}
	\item[a] Sum of all the good time intervals.
	\item[b] Those marked with ``*'' are  used in \citet{Cappi16}.
	\item[c] Simultaneous observation by \xmmnus. 
	\end{tablenotes}
	\end{threeparttable}
\end{table*}

\ngc is a typical well-studied Seyfert 1 galaxy, known to exhibit complicated X-ray spectral variations. 
It has complex and variable absorption features, and thus deep and multi-wavelength observation campaigns have been performed to reveal the surrounding structure around the central black hole.
The ``Anatomy of the AGN in NGC 5548'' campaign covered the wide wavelength range from near-infrared to hard X-rays during 2013--2014 (e.g., \citealt{Kaastra14}) and revealed the change of the absorbers' structure and geometry.
The ``Space Telescope and Optical Reverberation Mapping Project'' is a six-month-long reverberation-mapping experiment in 2014, which provided us with a deep insight into the broad line region (BLR) structure (e.g., \citealt{DeRosa15}).
In the ``Anatomy'' campaign, an outflowing stream of weakly ionized gas (called the ``obscurer''), which had not been present until at least 2007, was recognized, causing the soft X-ray absorption and the broad UV absorption \citep{Kaastra14}.
\citet{Kaastra14} and \citet{Arav15} also reported by spectral fitting of the Hubble Space Telescope UV data and \xmm X-ray data that the warm absorber (WA) is composed of six-velocity components.
 Using the electron number densities and the ionization parameters, \citet{Arav15} estimated  locations of these components as a few pc to a few hundred pc from the central black hole.

\ngc is one of the ideal sources to study gas structures around the AGN, thanks to its spectral variations and existence of the obscurer and the WA.
For the ``obscured'' \xmm and \nus energy spectra during the Anatomy campaign, 
\citet{Cappi16} applied a two-zone partial covering model with {\it different} covering fractions as well as other Anatomy campaign papers (e.g. \citealp{Ursini15}, \citealp{DiGesu15}, and \citealp{Mehdipour16}). \citet{Cappi16} explained the spectral variation in 0.3--50~keV with the following nine free parameters; (1)--(6) column densities, partial covering fractions, and ionization parameters of the two partial absorbers, (7)--(8) power-law photon index and the normalization, and (9) the soft excess normalization. Interestingly, they reported a correlation between one of the covering fractions and the photon index of the power-law continuum, such that the power-law slope gets steeper while the covering fraction increases. Since the covering fraction is determined by the geometry of the distant partial absorbers from the central black hole, this correlation may not be explained physically. 
Although \citet{Cappi16} mentioned the possibility of a parameter degeneracy, they  proposed a physical interpretation:
Assuming a disk-corona geometry similar to the one expected for Mrk~509 \citep{Petrucci13},  increase in the accretion rate might decrease the  inner radius of the corona due to disk recondensation (e.g. \citealp{Meyer-Hofmeister06}),  raising the effective covering fraction. 
Meanwhile,  the accretion rate increase produces more UV photons, causing more effective cooling in the corona and softer spectra.
Thus, a correlation between the covering fraction and the spectral index might be reproduced.

We suspect that the apparent correlation found by \citet{Cappi16} is more likely due to parameter degeneracy between the photon index and the mildly-ionized partial absorber. In fact,
when fitting a given CCD spectrum with increasing the partial covering fraction, the model soft X-ray photons are so decreased that the photon index has to get larger to compensate. 
To examine the possibility of parameter degeneracy, we need to constrain these parameters a priori in some manner.
Here, we try the ``variable double partial covering (VDPC) model'' \citep{Miyakawa12, Mizumoto14} expecting to solve this problem.
In this model, the line of sight is partially covered by clumpy absorbers composed of {\it double layers}; a thick/cold core and a thin/warm layer. 
Even if the intrinsic photon index is invariable, change of the partial covering fraction causes apparent spectral variations below $\sim$10~keV.
In this paper, we examine if the VDPC model can explain the seemingly complicated 2013--2014 energy spectra of \ngc.
Also, we apply the VDPC model to the X-ray spectra taken from 2000 to 2016 by \xmm, \su, and \nus to study the long-term spectral variations of \ngc.
 
Below, in Section~2, we introduce the observation and data reduction. Next, we explain  our spectral model and present the results of spectral fitting in Section~3. We also compare the observed Root Mean Square (RMS) spectra with the ones reproduced with the best-fit spectral model.
In Section~4, we discuss  application of the VDPC model to other objects,  origin of the multi-component absorbers, and the long-term X-ray variation of \ngc. 

\begin{figure*}
\centerline{\includegraphics[bb=150 12 750 360, width=1.2\columnwidth]{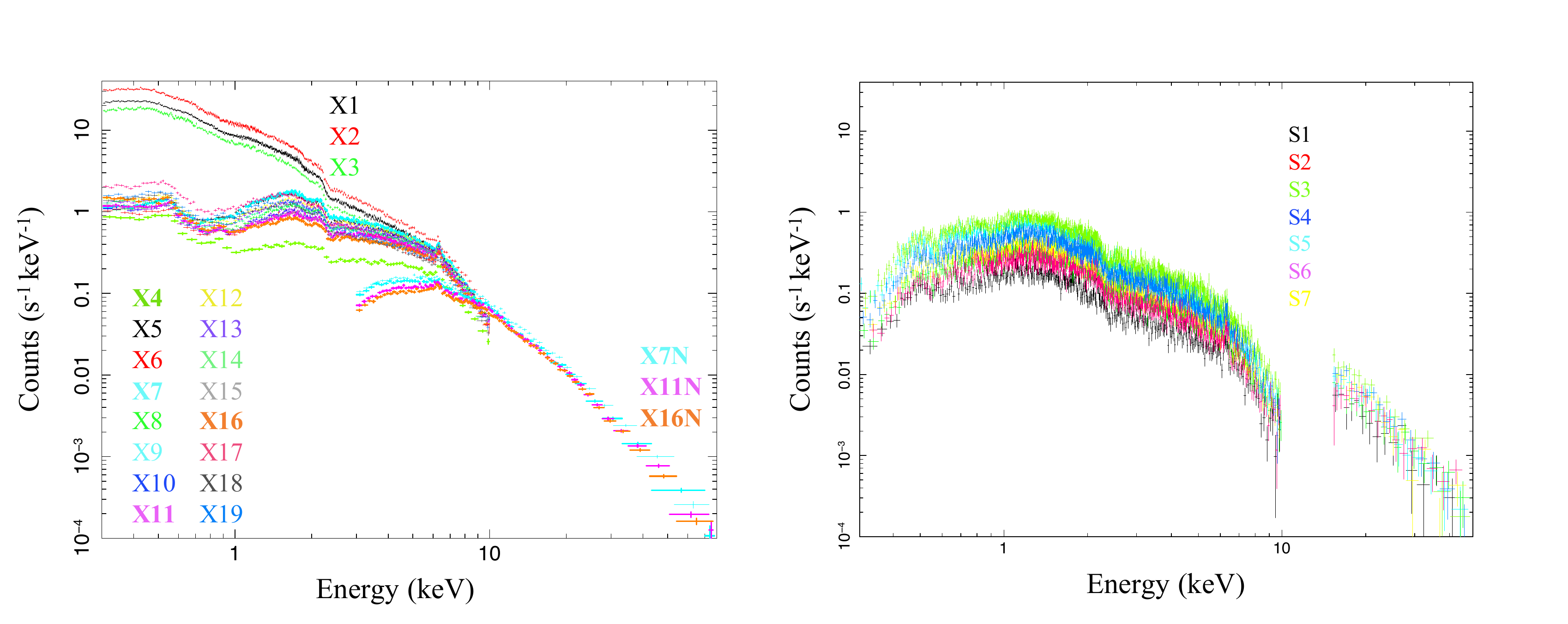}}
	    \caption{{\it Left}: All the archival spectra obtained by \xmm pn in 0.3--10.0~keV and by \nus in 3.0--78.0~keV.  
	    {\it Right}: All the archival spectra obtained by \su XIS1 in 0.3--10.0~keV and by PIN in 15.0--50.0~keV. Detector responses are removed using the best-fit models. 
	    }
	    \label{fig:all}
\end{figure*}

\section{Observations and Data Reduction}
We used all the available archival data taken by \xmm \citep{Jansen01a}, \nus \citep{Harrison13}, and \su \citep{Mitsuda07} during 2000--2016, including three simultaneous observations by \xmm and \nus (Table~\ref{tab:obs_table}).

In the analysis of the \xmm data, we used the European Photon Imaging Camera (EPIC)-pn \citep{Struder01a} in 0.3--10.0~keV and the Reflection Grating Spectrometer (RGS: \citealp{denHerder01}) in 0.4--2.0~keV.
The EPIC-MOS data were not used because its effective area is lower than that of the pn detector.
The pn and RGS data were reduced with SAS version 17.0.0 to obtain the filtered event files.
Good time intervals were selected by removing the intervals dominated by flaring particle background when the count rate in 10--12~keV with {\tt PATTERN==0} is larger than 0.4~counts~s$^{-1}$ for EPIC-pn data.
We used circular regions of 30\arcsec~radius and 45\arcsec~radius from the same CCD for extracting the source events and the background events, respectively.
Following the ``SAS Data Analysis Threads''\footnote
{\samepage{https://www.cosmos.esa.int/web/xmm-newton/sas-thread-epic-filterbackground\\https://www.cosmos.esa.int/web/xmm-newton/sas-thread-pn-spectrum\\https://www.cosmos.esa.int/web/xmm-newton/sas-thread-rgs}}, we obtained the X-ray spectra and responses.
The 1st-order spectra of RGS1 and RGS2 were combined.

To reduce the \nus data, following the ``\nus Data Analysis Software Guide''\footnote{https://heasarc.gsfc.nasa.gov/docs/nustar/analysis/nustar\_swguide.pdf}, we used the standard pipeline {\tt nupipeline} for reprocessing the data.
The source events were extracted from circular regions within a 1.64\arcmin ~radius around the sources, and the background events were extracted from an annulus region of 2.87\arcmin--4.10\arcmin ~around the source. 
The spectra were extracted from the cleaned event files using the standard tool {\tt nuproducts} for each of the two Focal Plane Modules (FPMA and FPMB).
The FPMA and FPMB spectra were combined to provide better statistics using {\tt addascaspec}. 

In the \su data analysis, we used the X-ray Imaging Spectrometer (XIS; \citealp{Koyama07}) 0, 1, and 3 in 0.3--10.0~keV and Hard X-ray Detector (HXD; \citealp{Takahashi07}).
The HXD consists of two types of detectors, PIN and GSO, achieving a combined sensitivity in 10--600~keV. We used only the PIN detector in this study (15--50~keV) because our target is not bright enough in the GSO energy band.
In the reduction of the XIS data, the source events were extracted from circular regions within a 2.0\arcmin ~radius of the sources, and the background regions were extracted from an annulus region of 3.0\arcmin--7.0\arcmin ~around the source. 
Spectra of the front-illuminated (FI) camera (XIS0 and XIS3) were combined by {\tt addascaspec}.
We adopted the ``tuned background'' as non X-ray background model of HXD-PIN\footnote{https://heasarc.gsfc.nasa.gov/docs/suzaku/analysis/pinbgd.html}.

We grouped {\textit{XMM}}-pn, {\textit{Suzaku}}-XIS, and \nus spectral bins so that each new bin holds more than 30 counts; similarly, 50 counts for {\textit{XMM}}-RGS spectra and 500 counts for {\textit{Suzaku}}-PIN spectra. 
Figure~\ref{fig:all} shows the 0.3--78.0~keV spectra of all the currently available archival data with \xmm, \nus, and \su.
In this research, we used HEASOFT version 6.26 and XSPEC version 12.10.1f for spectral model fitting.

\section{Data Analysis \& Results}

Our primary goal is to explain the apparently complex spectral variations of \ngc by a simple model with a minimum number of free parameters without parameter degeneracy.
In Section~\ref{sec:3.1}, we reproduce the correlation between the power-law index and one of the covering fractions   using the same two-zone partial covering  model by \citet{Cappi16}, and study whether the degeneracy disappears letting the two covering fractions tied.
In Section~\ref{sec:3.2}, we explain our VDPC model used in this study.
We first apply the VDPC model to the three \xmmnus simultaneous spectra and the dimmest \xmm spectrum to constrain the major parameters to characterize the spectral shape.
We  find these parameters are not significantly variable throughout the 16 year observation period.
In Section~\ref{sec:3.3}, we apply these invariant parameters to all the archival data and extract variations of other spectral parameters.

\begin{figure}
	\includegraphics[width=1.0\columnwidth]{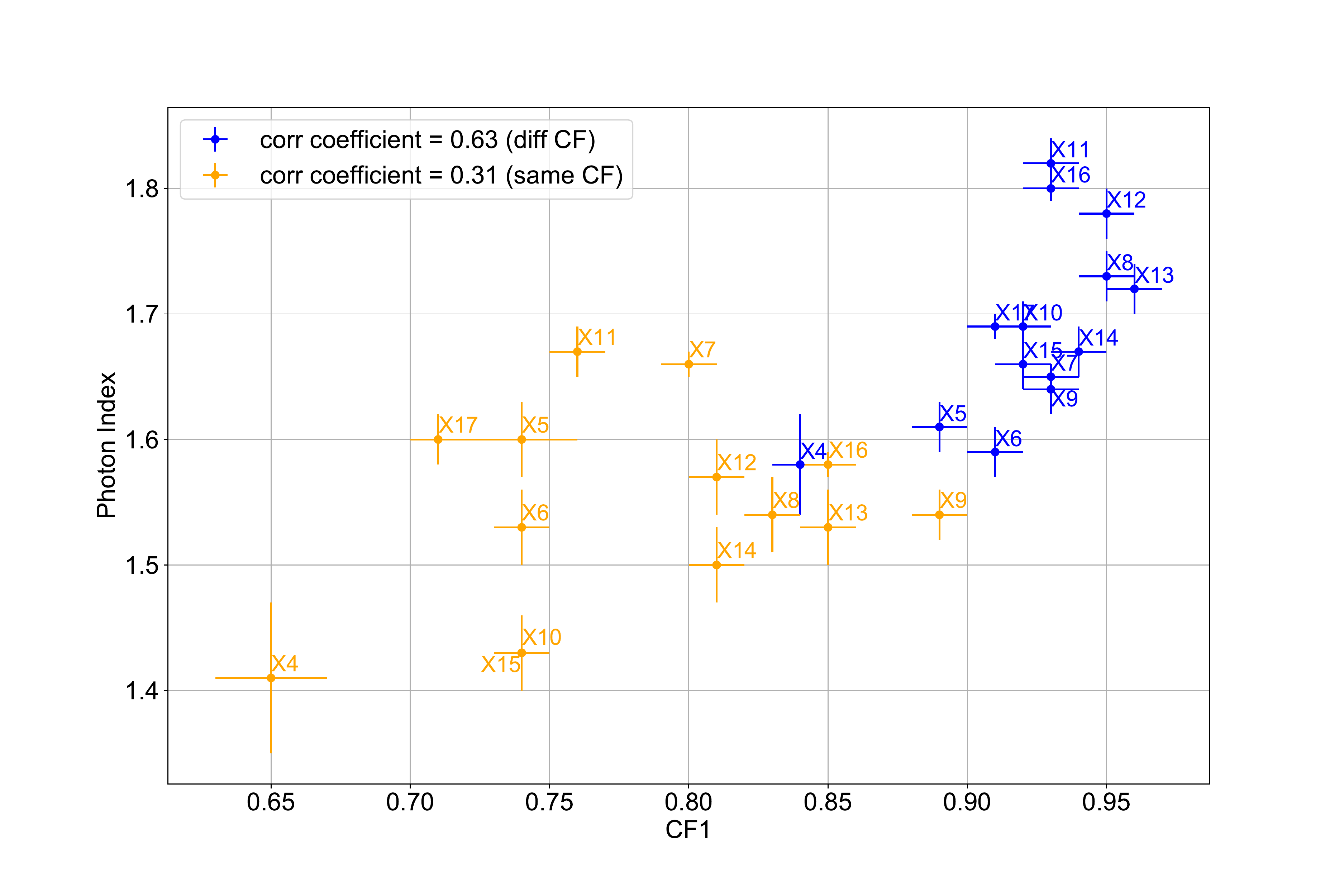}
	    \caption{Photon index versus one of the two covering fractions CF1. Blue/orange markers show the fitting results obtained when the two covering fractions are independent/tied. 
	    } 
	    \label{fig:Cappicorr}
\end{figure}

\subsection{Reproduction of the parameter correlation by Cappi et al. (2016)}\label{sec:3.1}
First, we examine the   \xmm spectral dataset used by \citet{Cappi16}.  
Using the same two-zone partial covering model by \citet{Cappi16}, we fitted allowing the two covering fractions to vary independently.
    We used XSTAR version 2.2.1bn21 \citep{Kallman04} for the plasma simulation code, instead of the Cloudy code \citep{Ferland13} adopted by \citet{Cappi16}. 
  As expected, we confirmed  the reported correlation between one of the partial covering fractions (CF1) and the photon index, where the
correlation coefficient is 0.63 (Figure~\ref{fig:Cappicorr}, blue crosses).
Next, we fitted 
the same dataset forcing  the two covering fractions to be identical, then the correlation coefficient decreased to 0.31 (Figure~\ref{fig:Cappicorr}, orange crosses).   If we ignore an outlier (X4), the coefficient further drops to  0.08.  
Consequently, we suggest that the reported correlation is most likely due to the parameter degeneracy, and reduction of the free parameters can explain the spectral variation without any unphysical correlations.

\subsection{Spectral fitting to the simultaneous  \xmmnus spectra}\label{sec:3.2}
We apply our VDPC model to the spectral data.
First, we select  the \xmmnus simultaneous spectra (X7, X7N, X11, X11N, X16, X16N in Table~\ref{tab:obs_table}) and the dimmest \xmm spectrum (X4) to constrain the main spectral parameters. 
Since both the power-law (\pl) and the reflection components (\pexmon) are dominant above $\sim$10~keV, and effect of the partial absorption is most prominent in the dimmest spectrum, we expect to effectively constrain the overall spectral shape by fitting these spectra simultaneously.

The VDPC model adopted in this work is represented by the following equation:
\begin{eqnarray}
\begin{split}
    F =& ~[(P+B)(1-\alpha+\alpha W_1)(1-\alpha+\alpha W_2) W_3 W_4 + R_{\rm I_{Fe}} + I_{{\rm gas}}]A\\
    =& ~[(P+B)\{(1-\alpha)(1-\alpha) + \alpha W_1 (1-\alpha) + (1-\alpha)\alpha W_2\\
    & + \alpha ^2 W_1 W_2)\} * W_3 W_4 + R_{\rm I_{Fe}} + I_{{\rm gas}}]A, 
\label{eq:vdpc}
\end{split}
\end{eqnarray}
where meaning of the parameters are explained below.
The XSPEC representation is written as:
\begin{eqnarray}
\begin{split}
    F =&~{\tt \{(cutoffPL+diskbb)(1-const+const*mtable_1)}\\
    &~*{\tt (1-const+const*mtable_2)*mtable_3*mtable_4}\\ 
    &~{\tt + pexmon + mekal\}*phabs}.
\label{eq:xspec}
\end{split}
\end{eqnarray}

\begin{figure}
\includegraphics[width=1.0\columnwidth]{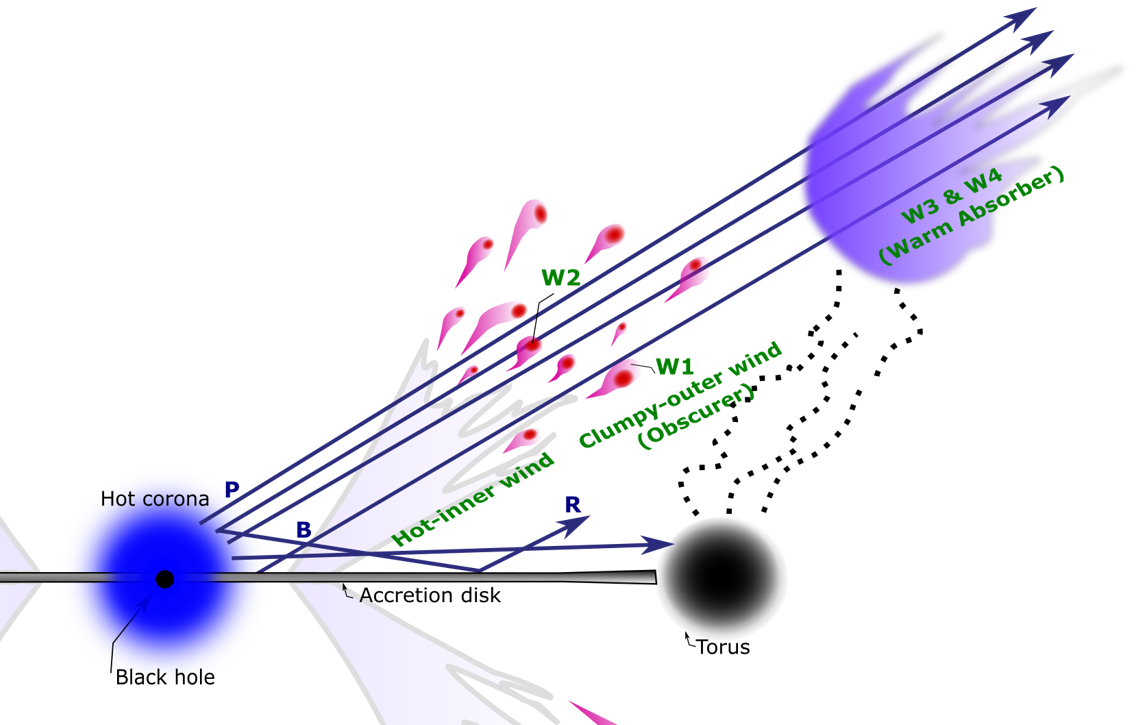}
	    \caption{Schematic view of our picture around the black hole and accretion disk (not in scale). The intrinsic X-rays of \pl ($P$ in the figure) and \diskbb ($B$) are emitted from the hot corona around the central black hole and the inner accretion disk, respectively. These intrinsic X-rays are partially covered by the ``cold clumps'', each of which consists of two layers; a cold core ($W_2$) and warm layer ($W_1$). These cold clumps,  known as ``obscurers'', cause the ``double partial covering'' with the same covering fraction. The cold clumpy-outer wind is likely to be a consequence of the putative hot-inner wind that is hardly recognized probably because too hot and/or optically thin. All the X-rays are fully absorbed by the outer WAs ($W_3$ and $W_4$), which are formed from the ionized material due to evaporation of the irradiated torus.}
	    \label{fig:otama}
\end{figure}

A cut-off power-law component from the hot corona and the \diskbb component from the optically thick accretion disk are the direct X-ray continua (see Figure~\ref{fig:otama}). 
We suppose that the clumpy absorber has an internal double-layer structure that consists of a warm layer ($W_1$) and a cold core ($W_2$).
These clumpy absorbers partially obscure the intrinsic continuum, where changes of the partial covering fractions in the line of sight are causing the apparent spectral variations. 
Then X-rays are fully absorbed by the ionized absorbers ($W_3$ and $W_4$) located farther from the center than the partial absorbers.
Moreover, the power-law component is reflected by the outer accretion disk accompanying a narrow Fe emission line (\pexmon).
Besides these model components, there is a constant plasma thermal emission (\mekal) that accounts for the emission lines in the soft X-ray band.

The cosmic redshift is 0.017 \citep{Cappi16}, and we set the cosmic abundances to the {\tt wilms} values \citep{Wilms00a}. 
Since the spectral slopes above $\sim$10~keV looks very similar (Figure~\ref{fig:all}, left), we adopt a simple cut-off power-law model where the index and the cut-off energy are invariable.
Cross-normalization constants between the three instruments ({\textit{XMM}}-pn, {\textit{XMM}}-RGS, and {\textit{NuSTAR}}-FPM) were  determined by the simultaneous fitting, resulting in the consistent values  with \citet{Madsen17}.  

\begin{figure}
	\includegraphics[width=1.0\columnwidth]{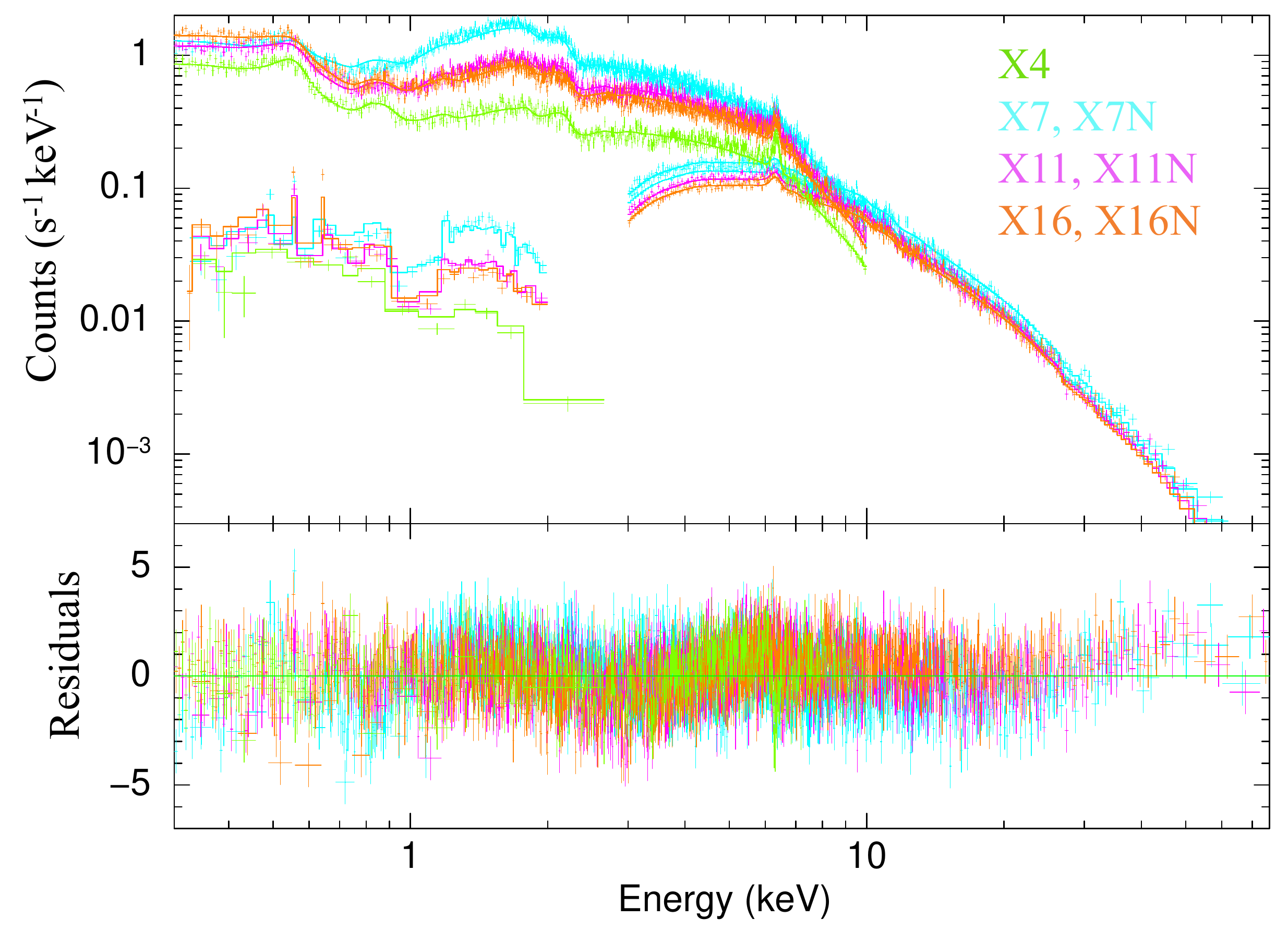}
	    \caption{Spectral fitting of the three simultaneous observations by \xmmnus (X7, X11, X16) and the dimmest observation by \xmm (X4) within 0.3--78.0~keV. The upper panel shows the observed spectra and the best-fit model (see table~\ref{tab:invar_par}). 
	    The lower panel shows  residuals of the model fitting.}
	    \label{fig:4obs}
\end{figure}

We have tried to  fit  the seven spectra of the four observations  simultaneously with minimum free parameters.  Consequently, 
we have found that these observations  can be explained without  varying the power-law
index significantly, and the reflection component and the multiple absorber parameters at all (Table \ref{tab:invar_par} and Figure \ref{fig:4obs}). 
The power-law photon-index can be the same at 1.54, besides the dimmest observation (X4), 1.39.
In Table~\ref{tab:invar_par}, we show the  best-fit parameters  common to the four observations. 
Other variable parameters are shown in Table~\ref{tab:var_par}.
Figure~\ref{fig:4obs} shows the results of the fitting in 0.3--78.0~keV, where the reduced chi-square value is 1.24 (d.o.f. $=$ 9418). 

The following is  explanation of the individual model components in equations 
(\ref{eq:vdpc}) and (\ref{eq:xspec}):

\begin{enumerate}
    \item {\it Interstellar absorption}: $A$/\phabs: $\exp(-\nh \sigma(E))$\\
\nh and $\sigma(E)$ are the hydrogen column density along the line of sight and the photoelectric absorption cross section as a function of the X-ray energy, respectively. The column density was fixed at the Galactic value \nh $= 1.55\times 10^{20}$ cm$^{-2}$ \citep{Bekhti16}. 
    \item {\it Cut-off power-law}: $P$/\pl: $KE^{-\Gamma} \exp(-E/E_{\rm cut})$\\
$K$ is the normalization in [photons/keV/cm$^2$] at 1~keV, $\Gamma$ is the photon index and $E_{\rm cut}$ is the cut-off energy. We assume that the power-law component is emitted from the hot corona around the central black hole (Figure~\ref{fig:otama}).
    \item {\it Partial absorption by ``cold clumps''}: $W_{1,2}$/$\mtable_{1,2}$: $\exp\{-\sigma(E, \xi)N_{\rm H}\}$\\
These are the essential components in the VDPC model and correspond to the ``obscurer'' recognized in the ``Anatomy'' campaign observation \citep{Kaastra14}.
The {\tt mtable} models describe transmission by ionized absorbers as a function of the column density $N_{\rm H}$ and the ionization parameter $\xi$, calculated with XSTAR.
We adopt the 20$\times$20 parameter grids in the range of  0.1 < log  $(\xi$~/~erg~cm~s$^{-1})$ < 5 and 20 < log $(N_{\rm H}$~/~cm$^{-2}$) < 25 in logarithmic steps.
We fix the gas temperature at 10$^5$~K, the gas density at $1.1\times 10^{12}$~cm$^{-3}$, and the turbulent velocity at 200~km~s$^{-1}$.
Abundances are fixed to the solar values except for those  minor elements (Be, Ba, Na, Al, P, Cl, K, Sc, Ti, V, Cr, Mn, Co, Cu, and Zn) that are ignored to simplify the  calculation.
The incident spectrum is assumed to be a power-law with the photon index  2.0. Effects of changing the intrinsic spectra  are estimated in the final paragraph of this section.

Combination of the fixed \mtable and the variable covering fraction $\alpha$/{\tt const} represents the partial absorption. We need two different ionized partial absorbers, which have the same $\alpha$. 
These partial absorbers are presumably due to double-layered cold clumps (Figure~\ref{fig:otama}). 
We will discuss physical interpretation of these absorbers in detail in Section~\ref{sec:4.1}.
    \item {\it Full absorption by ``ionized absorber''}: $W_{3,4}$/$\mtable_{3,4}$: $\exp\{-\sigma(E,\xi)N_{\rm H}\}$\\
These two components cause the ionized full-covering absorption  (Figure~\ref{fig:otama}), which approximate several components of WAs. 
\citet{Cappi16} noted that the six-component WAs reported in \citet{Kaastra14} are  approximated by two components  during the unobscured period and that  influence of the multi-components is weaker during the obscured period.
We applied  average values of the two WAs by \citet{Cappi16}
to the invariable components $W_3$ and $W_4$.
    \item {\it Cold reflection with Fe line}: $R_{\rm I_{Fe}}$/\pexmon\\
This is a Compton reflection component from cold outer-part of the accretion disk, including the narrow fluorescence lines (\pexmon; \citealp{Nandra07}). We fixed the disk inclination angle at $i = 30^\circ$, where it was suggested as $i=38.8^{+12.1}_{-11.4}$\relax$^{\circ}$ \citep{Pancoast14} or $36\pm10^\circ$ \citep{Starkey17}.
We determined the redshift value by the simultaneous fitting considering the line shift from the cosmic value due to an ionized disk or a kinetic Doppler shift.
Since the observed fluorescent Fe K${\rm \alpha}$ line strength does not vary significantly compared to the continuum variation, we have fixed the \pexmon normalization in the simultaneous fitting.
    \item {\it Soft excess}: $B$/\diskbb\\
Several interpretations exist for the soft excess component, e.g., the optically thick Comptonized disk model \citep{Done12}.
In this paper, we assume a  black body accretion disk model  for simplicity (\diskbb; \citealp{Mitsuda84}; \citealp{Makishima86}).
The \diskbb has two parameters, temperature at the inner edge of the accretion disk $kT_{\rm in}$ and normalization ($=r_{\rm in}^2 \cos{i}/d_{\rm 10kpc}$), where $r_{\rm in}$, $d_{\rm 10kpc}$, and $i$ are the innermost radius of the disk, the distance in the unit of 10~kpc, and the disk inclination angle, respectively.  In  most cases of \ngc, only the soft excess tail  is observed below $\sim$1~keV, and the temperature and the normalization are hardly constrained simultaneously. Thus, we fixed the \diskbb normalization, and 
varied only $kT_{\rm in}$.
    \item {\it Thermal plasma emission}: $I_{\rm gas}$/\mekal\\
The soft excess component has not only the continuum emission but also the emission lines.
We modeled the emission-line component using the thermal plasma radiation model \mekal. Since one-temperature \mekal does not accurately describe the observed spectrum, we apply two-temperature \mekal in this study. This component has been constant throughout the 16-year observation. 
The hot plasma is considered to  originate either in  the host galaxy or in  the narrow line region (e.g., \citealp{Whewell15}; \citealp{Cappi16}).
\end{enumerate}

In the actual multi-absorber  scenario, we note that  the first absorber would alter the input  spectrum to the second one, and so does the second one to the  input spectrum to the third one, and so on. Thus,
it would be ideal to carry out such successive photo-ionization calculation that a preceding absorber's output spectrum is  incident to the following absorber. However, this is not so easy that we have adopted such a simple approximation that the incident spectrum to
the absorbers is a static power-law function.

Below, we evaluate the validity of this approximation.
We investigate whether the spectral fitting results
are dependent on XSTAR table models ({\tt mtable}) created with different incident spectra.
Table~\ref{tab:invar_par} shows the best-fit parameters using the table model created with the photon index 2.0.  We created new
table models with the input photon indices 1.5 and 2.5, and repeated  the same  simultaneous fitting
replacing  the tables.  Consequently, we found that the fitting results are almost identical (the reduced chi-square values are  1.26 and 1.22, respectively, as opposed to the original one, 1.24),  and the residual patterns are not distinguishable. 
Best-fit  parameter values
are invariable within statistical uncertainty, besides those of the partial absorber parameters (\nh and $\log \xi$ for  $W_1$ and \nh for $W_2$). The new best-fit values are the following (in the order of the  photon-indices 1.5 and 2.5):  For  $W_1$,  (\nh~/~10$^{22}$cm$^{-2}$)=1.17 and 1.05,  log ($\xi$~/~erg~cm~s$^{-1}$)=0.93 and 0.1. 
For $W_2$, (\nh~/~10$^{22}$cm$^{-2}$)=2.46 and 3.04.  These differences are minor, and  we conclude that our approximation of the constant  incident spectra to the multi-absorbers is valid to represent the picture depicted in Figure \ref{fig:otama}.

\begin{table} 
	\begin{center}
	\caption{Spectral parameters determined by simultaneous fitting  in 0.3--78.0~keV of the seven spectra from  three simultaneous observations by \xmmnus (X7, X11, X16) and the dimmest observation by \xmm (X4). Model details are explained in Section~\ref{sec:3.2}. The parameters shown in this table are common in the four observations, X4, 7, 11, and 16, except for \pl $\Gamma$. The photon index of X4 is smaller than those of the other observations and cannot be fitted well with the tied value. The error ranges correspond to the 90\% confidence level throughout this paper.}
	\label{tab:invar_par}
    \begin{threeparttable}
    \begin{tabular}{cccc}\hline
        & Model & Parameters & Best-fit values \\\hline
        $A$ & \phabs & \nh (10$^{20}$cm$^{-2}$) & 1.55 (fixed)  \\\cdashline{1-4}[3pt/1.5pt]
        $W_1$ & warm-layer absorber & \nh (10$^{22}$cm$^{-2}$) & 1.10$\pm0.05$  \\
         &  & log $\xi$ & 0.40$\pm0.02$  \\\cdashline{1-4}[3pt/1.5pt]
        $W_2$ & cold-core absorber & \nh (10$^{22}$cm$^{-2}$) & 2.70$\pm0.04$  \\
         &  & log $\xi$ & 0.10 (fixed) \\\cdashline{1-4}[3pt/1.5pt]
        $W_3$ & ionized absorber & \nh (10$^{22}$cm$^{-2}$) & 0.24 (fixed)  \\
         &  & log $\xi$ & 1.07 (fixed)  \\\cdashline{1-4}[3pt/1.5pt]
        $W_4$ & ionized absorber & \nh (10$^{22}$cm$^{-2}$) & 0.66 (fixed)  \\
         &  & log $\xi$ & 2.70 (fixed)  \\\cdashline{1-4}[3pt/1.5pt]
        $P$ & \pl & $\Gamma$ (X7,X11,X16) & 1.54$^{+0.01}_{-0.02}$  \\
         &  & $\Gamma$ (X4) & 1.39$\pm0.02$  \\
         &  & E$_{\rm cut}$(keV) & 72.8$^{+5.3}_{-5.0}$  \\\cdashline{1-4}[3pt/1.5pt]
        $R_{\rm I_{Fe}}$ & \pexmon & $\Gamma$ & $=$ \pl $\Gamma$ \\
         &  & abund & 1.0 (fixed) \\ 
         &  & redshift & 0.009$\pm0.001$  \\ 
         &  & norm (10$^{-3}$) & 2.5$\pm0.2$ \\\cdashline{1-4}[3pt/1.5pt]
        $B$ & \diskbb & norm (10$^{3}$) & 44.0$^{+7.5}_{-6.5}$ \\\cdashline{1-4}[3pt/1.5pt]
        $I_{\rm gas}$\tnote{a} & {\tt mekal$_1$} & $kT$ (keV) & 0.142$\pm0.008$  \\
         &  & norm (10$^{-4}$) & 3.3$\pm0.6$  \\
         & {\tt mekal$_2$} & $kT$ (keV) & 0.623$\pm0.026$  \\
         &  & norm (10$^{-4}$) & 0.82$\pm0.08$  \\\hline
	\end{tabular}
	\begin{tablenotes}
    	\item[a] The \mekal parameters were determined only from X4  since they are hardly constrained in other spectra due to confusion of  other spectral components.
	\end{tablenotes}
	\end{threeparttable}
	\end{center}
\end{table} 

\subsection{Fitting of all the available spectra}\label{sec:3.3}
We have shown in the previous section that the four 
spectra in the 2013 campaign (X4, X7, X11, and X16) including the brightest (X7) and the dimmest (X4) ones are explained without varying the main spectral shape and the multiple absorber parameters (Table \ref{tab:invar_par} and Figure \ref{fig:4obs}). Most variable
parameters are  (1) the partial covering fraction, (2) the power-law normalization and
(3) the \diskbb temperature (Table \ref{tab:var_par}). We examine if similar spectral fitting is possible  for 
all the other observations with minimum numbers of the variable parameters.

We fit each of all the observations individually 
 with minimum variable parameters in addition to  the three free-parameters in the 2013 \xmmnus campaign. 
 First,  we found no significant variations in the iron emission line and the reflection component (\pexmon).
We also found that the spectra taken during the unobscured period (X1, X2, and X3 in 2000--2001)
had clearly different characteristics from those taken in 2007 and afterward.  
These unobscured spectra were much brighter than others below $\sim$2~keV  (Figure \ref{fig:all}),
where the covering fraction was null.  They  required a higher photon index and a lower soft excess normalization than the 2013 \xmmnus campaign.
These results suggest that the physical conditions of the hot corona and the inner accretion disk  had 
significantly varied in the  time-scale of years between 2001 and 2007, whereas the  outer disk
reflection is invariable in this time-scale.  

For all the other spectra  
 in 2007 and afterward, 
the same  index (1.54) as in the simultaneous fitting  is valid  beside the faintest spectrum  X4 which requires a lower photon-index (1.39).
The \diskbb normalization was also fixed to the value determined from the \xmmnus simultaneous fitting.
We show all the results of the model fitting in Figure~\ref{fig:allxmm_fit} (for \su) and Figure~\ref{fig:allsu_fit} (for \xmm). 
The best-fit parameters are shown in Table~\ref{tab:var_par}.
Some spectra (e.g., X5, X15 and X17) exhibit  local residual feature below $\sim$1 keV, which would suggest a minor variation of the absorbers.
Besides, major variable parameters over the entire $\sim$16 year period are (1) the covering fraction $\alpha$, (2) the \pl normalization, and (3) the \diskbb temperature $kT_{\rm in}$, as in the 2013 \xmmnus campaign. 
We emphasize that all the model fits are reasonable  with such a  simple model
with little changes in the intrinsic spectra and absorbers' parameters.

We point out that 
the observation X4, which was the faintest in the \xmm observations,  returns a photon index of 1.39, that seems extremely hard.  Such low photon-indices in extreme low states have 
been also reported in other Seyfert galaxies
\citep[e.g.,][]{2004ApJ...605..670P}.  It is possible that 
the the AGN corona may have a tendency to get hotter and/or thicker when getting dimmer.

During the \su observation period (June to August 2007), the covering fraction was tiny ($<$ 0.05) or null, which may imply that the obscurer started to emerge at around that time.
While previous studies (e.g., \citealp{Liu10}; \citealp{Krongold10}) analyzing the same \su data have not reported the presence of the obscurer, \cite{Kaastra14} argued that the obscuration may have started between August 2007 and February 2012, when {\it Swift}\/ was not monitoring \ngc.

\begin{table*}
\begin{center}
\caption{Variable parameters determined by the spectral fitting with \xmm (pn, RGS) and \su (XIS0, 1, 3, PIN). Other parameters common to all the spectra are shown in Table~\ref{tab:invar_par}. Those parameters without error values are the fixed ones.}
\label{tab:var_par}
\scalebox{0.95}{
\begin{threeparttable}
\tabcolsep = 4pt
\begin{tabular}[H]{|c|ccccc:c|cc|}\hline
 \multirow{2}{*}{Obs.} & \multicolumn{6}{c|}{fitting results (refer Section~\ref{sec:3.3})} & \multicolumn{2}{c|}{computed flux (refer Section~\ref{sec:4.2})}\\\cline{2-9} 
& $\alpha$ & \begin{tabular}{c}\diskbb \\$kT_{\rm in}$ (eV)\end{tabular} & \begin{tabular}{c}\diskbb \\norm ($10^3$)\end{tabular} & \begin{tabular}{c}PL norm\\ (10$^{-3}$)\end{tabular} & PL index & $\chi^2$/dof & \begin{tabular}{c}\diskbb flux\tnote{a}\\ ($10^{-12}$erg/s/cm$^2$)\end{tabular} & \begin{tabular}{c}PL flux\tnote{b}\\ ($10^{-12}$erg/s/cm$^2$)\end{tabular}\\\hline
X1 & 0 & 162$\pm3$ & 1.21$\pm0.09$ & 8.4$\pm0.1$ & 1.70 & 1.17 & 1.4$\pm0.1$ & 129$\pm1.5$ \\ 
X2 & 0 & 176$\pm1$ & 0.83$\pm0.03$ & 10.5$\pm0.1$ & 1.70 & 1.41 & 0.9$\pm0.1$ & 161$\pm1.5$ \\
X3 & 0 & 186$\pm2$ & 1.04$\pm0.05$ & 13.8$\pm0.1$ & 1.70 & 1.33 & 1.8$\pm0.1$ & 211$\pm1.5$ \\\hline
S1 & 0.02$^{+0.02}_{-0.01}$ & 57$^{+2}_{-3}$ & 44.0 & 1.6$\pm0.1$ & 1.54 & 1.03 & 27.8$\pm5.9$ & 24$\pm1.5$ \\
S2 & 0.02$\pm0.01$ & 50 & 44.0 & 2.8$\pm0.1$ & 1.54 & 1.06 & 16.4 & 43$\pm1.5$ \\
S3 & 0 & 72$\pm1$ & 44.0 & 6.0$\pm0.1$ & 1.54 & 1.31 & 70.7$\pm3.9$ & 92$\pm1.5$ \\
S4 & 0.04$\pm0.01$ & 62$\pm2$ & 44.0 & 3.7$\pm0.1$ & 1.54 & 1.06 & 38.9$\pm7.5$ & 57$\pm1.5$ \\
S5 & 0 & 71$\pm1$ & 44.0 & 7.4$\pm0.1$ & 1.54 & 1.10 & 66.8$\pm3.8$ & 113$\pm1.5$ \\ 
S6 & 0 & 55$^{+4}_{-9}$ & 44.0 & 4.7$\pm0.1$ & 1.54 & 1.06 & 24.1$\pm8.8$ & 72$\pm1.5$ \\
S7 & 0.05$\pm0.01$ & 55$^{+3}_{-5}$ & 44.0 & 2.5$\pm0.1$ & 1.54 & 1.08 & 24.1$\pm3.5$ & 38$\pm1.5$ \\\hline
X4 & 0.96$\pm0.01$ & 137$\pm1$ & 44.0 & 2.4$\pm0.1$ & 1.39$\pm0.01$ & 1.13 & 926.6$\pm27.1$ & 37$\pm1.5$ \\
X5 & 0.71$\pm0.01$ & 50 & 44.0 & 7.5$\pm0.1$ & 1.54 & 1.36 & 16.4 & 115$\pm1.5$ \\
X6 & 0.76$\pm0.01$ & 74$\pm1$ & 44.0 & 5.2$\pm0.1$ & 1.54 & 1.21 & 78.9$\pm4.3$ & 80$\pm1.5$ \\
X7 & 0.73$\pm0.01$ & 50 & 44.0 & 8.2$\pm0.1$ & 1.54 & 1.38 & 16.4 & 125$\pm1.5$ \\
X8 & 0.78$\pm0.01$ & 78$\pm1$ & 44.0 & 6.7$\pm0.1$ & 1.54 & 1.21 & 97.4$\pm5.0$ & 103$\pm1.5$ \\
X9 & 0.81$\pm0.01$ & 84$\pm1$ & 44.0 & 6.4$\pm0.1$ & 1.54 & 1.29 & 131.0$\pm6.2$ & 98$\pm1.5$ \\
X10 & 0.77$\pm0.01$ & 77$\pm1$ & 44.0 & 5.5$\pm0.1$ & 1.54 & 1.42 & 92.5$\pm4.8$ & 80$\pm1.5$ \\
X11 & 0.95$\pm0.01$ & 146$\pm1$ & 44.0 & 6.2$\pm0.1$ & 1.54 & 1.19 & 1195.1$\pm32.7$ & 95$\pm1.5$ \\
X12 & 0.74$\pm0.01$ & 78$\pm1$ & 44.0 & 7.3$\pm0.1$ & 1.54 & 1.32 & 97.4$\pm5.0$ & 112$\pm1.5$ \\
X13 & 0.75$\pm0.01$ & 78$\pm1$ & 44.0 & 7.2$\pm0.1$ & 1.54 & 1.25 & 97.4$\pm5.0$ & 110$\pm1.5$ \\
X14 & 0.74$\pm0.01$ & 78$\pm1$ & 44.0 & 6.4$\pm0.1$ & 1.54 & 1.31 & 97.4$\pm5.0$ & 98$\pm1.5$ \\
X15 & 0.77$\pm0.01$ & 80$\pm1$ & 44.0 & 5.6$\pm0.1$ & 1.54 & 1.48 & 107.7$\pm5.4$ & 86$\pm1.5$ \\
X16 & 0.94$\pm0.01$ & 142$\pm1$ & 44.0 & 5.4$\pm0.1$ & 1.54 & 1.28 & 1069.4$\pm30.1$ & 83$\pm1.5$ \\
X17 & 0.59$\pm0.01$ & 69$\pm1$ & 44.0 & 6.1$\pm0.1$ & 1.54 & 1.70 & 59.6$\pm3.5$ & 93$\pm1.5$ \\
X18 & 0.63$\pm0.01$ & 64$\pm1$ & 44.0 & 5.0$\pm0.1$ & 1.54 & 1.34 & 44.1$\pm2.8$ & 77$\pm1.5$ \\
X19 & 0.62$\pm0.01$ & 66$\pm1$ & 44.0 & 5.4$\pm0.1$ & 1.54 & 1.27 & 49.9$\pm3.0$ & 83$\pm1.5$\\\hline
\end{tabular}
	\begin{tablenotes}
	\item[a] Bolometric fluxes of \diskbb without any absorption.
	\item[b] \pl fluxes without any absorption in 0.3--50.0~keV assuming the best-fit parameters $\Gamma = 1.54$ and E$_{\rm cut} = 72.8$~keV. 
	\end{tablenotes}
	\end{threeparttable}}
    \end{center}
\end{table*}

\subsection{Reproduction of RMS spectra with the best-fit model}\label{sec:3.4}
In this section, we try to explain the observed RMS spectra with the best-fit spectral model obtained in the previous section.
The RMS spectra show fractional variations as a function of the X-ray energy, that intuitively informs us which  components primarily contribute to the spectral variability.

\begin{figure*}
	\includegraphics[width=2.05\columnwidth]{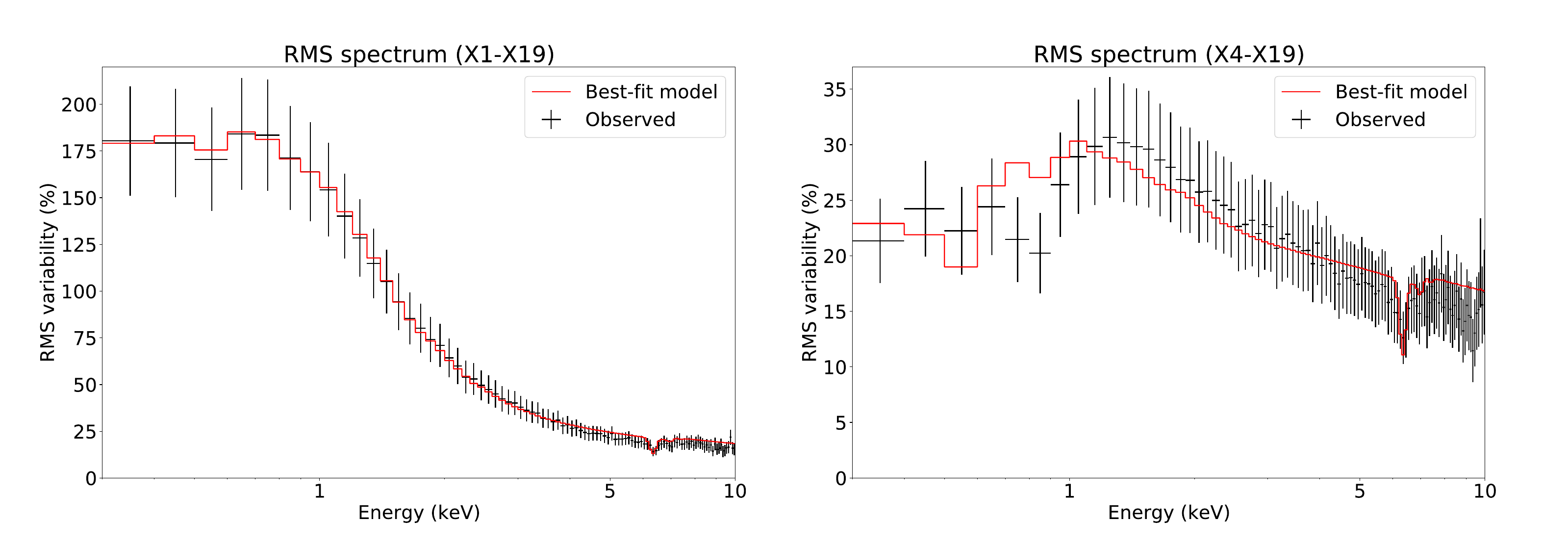}
	    \caption{RMS spectra calculated from the observed spectra (black) and the best-fit model spectra (red). Observations used for RMS calculation are X1--X19 (left) and only X4--X19 (right), respectively.}
	    \label{fig:rms}
\end{figure*}

Following \cite{Edelson02}, the RMS spectra were calculated from the observed and best-fit model spectra for each \xmm observation with the 0.1~keV energy bins between 0.3--10.0~keV.
Figure~\ref{fig:rms} shows the observed and model RMS spectra. The left one shows the RMS spectra from all the  observations including the unobscured period (X1--X3), and 
the right  is calculated from only X4--X19 in the obscured period.

In both figures, the dip around 6.4~keV indicates that the neutral Fe line from the outer-disk reflection is relatively constant compared to other spectral components.
In the left figure, the variations are as large as $\sim$200 \% below 1~keV, which is due to the emergence of the obscurers after 2007.
In the right figure, where most variations are due to change of the partial covering fraction,
the RMS spectrum is characterized by a broad peak around $\sim$1 keV.
This is because the partial absorber is  most opaque just below the Fe-L edge,
thus variation of the partial covering fraction most affects the spectral variation
at around $\sim$1 keV (see also Figure \ref{fig:setpladd}).
 Similar $\sim$1 keV peaks are recognized from RMS spectra of other Seyfert 1 galaxies, where partial covering
is  dominant \citep[e.g., ][]{Yamasaki16}.

\section{Discussion}

We have found that the VDPC model, where two partial absorbers have exactly the same covering fraction, can explain the long-term spectral variations and RMS spectra of \ngc.
This model is shown to be also valid for other Seyfert galaxies in precedent works \citep{Miyakawa12,Mizumoto14,Iso16,Yamasaki16}. 
Below, we discuss physical origin and implication of the VDPC model.

\subsection{Mathematical and Physical description of the VDPC model}\label{sec:4.1}
\begin{figure}
	\includegraphics[width=1\columnwidth]{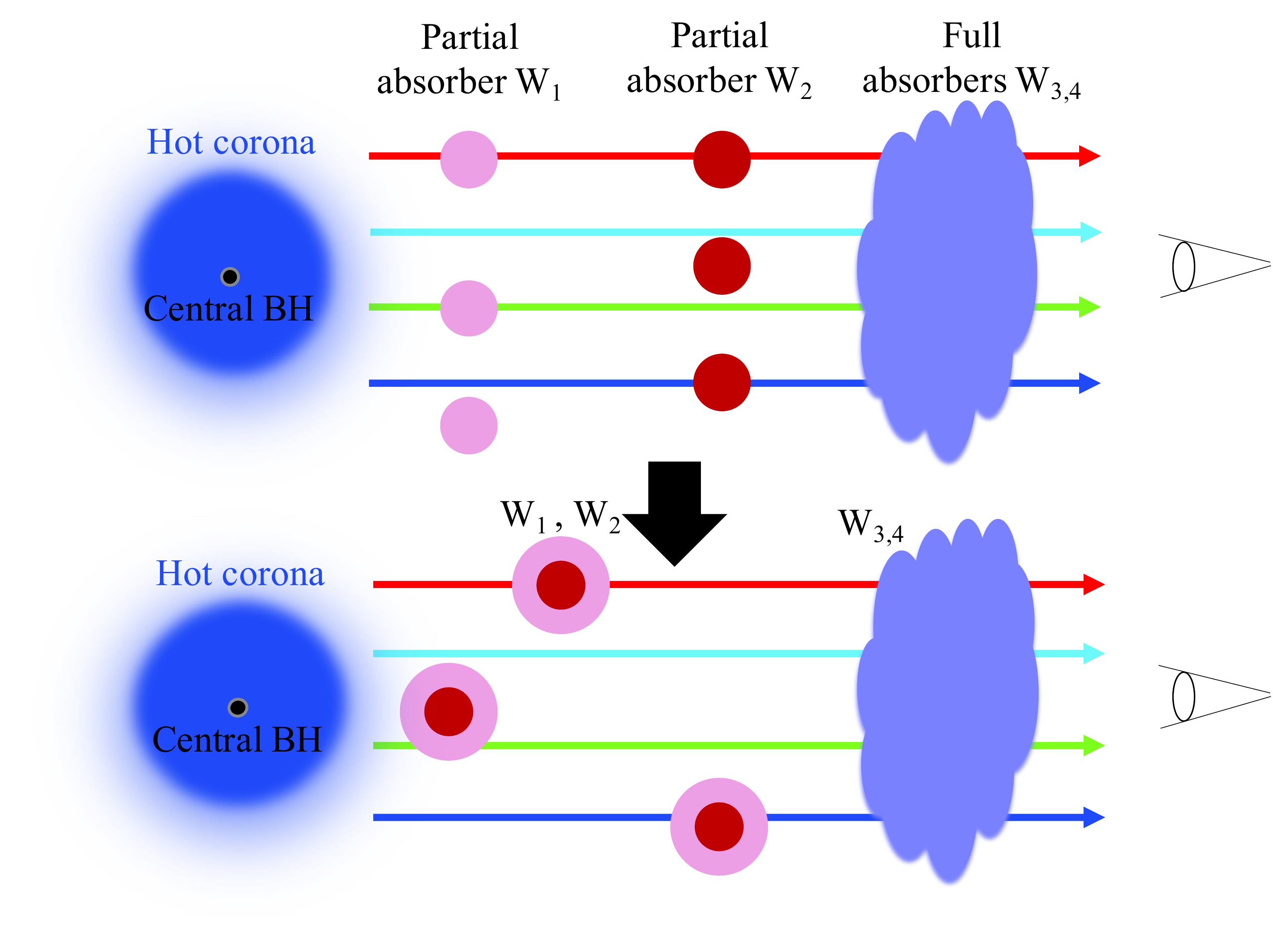}
	    \caption{Schematic drawings of the absorbers with the VDPC model. {\it Upper}\/: Faithful schematic description of the absorbers. The same fraction of the X-rays (in this case 50 \%) are absorbed by the two independent absorbing layers, $W_1$ and $W_2$. {\it Lower}\/ : More realistic configuration of the absorbers.}
	    \label{fig:schematic}
\end{figure}

The mathematical VDPC representation is described in Equation~\ref{eq:vdpc}, which indicates a situation that the central X-ray emitting region is sequentially obscured by the partial absorbers $W_1$ and $W_2$ having the same covering fraction and fully covered by the full-absorbers $W_3$ and $W_4$.
The upper diagram in Figure~\ref{fig:schematic} literally illustrates this situation.
The key point of the VDPC model is that the two partial absorbers $W_1$ and $W_2$ have exactly the same and {\em variable}\/ covering fractions.
However, it is very challenging to assume that the two independent absorbing layers accidentally have the same covering fraction and always vary in sync. 

\begin{figure*}
\includegraphics[width=2.05\columnwidth]{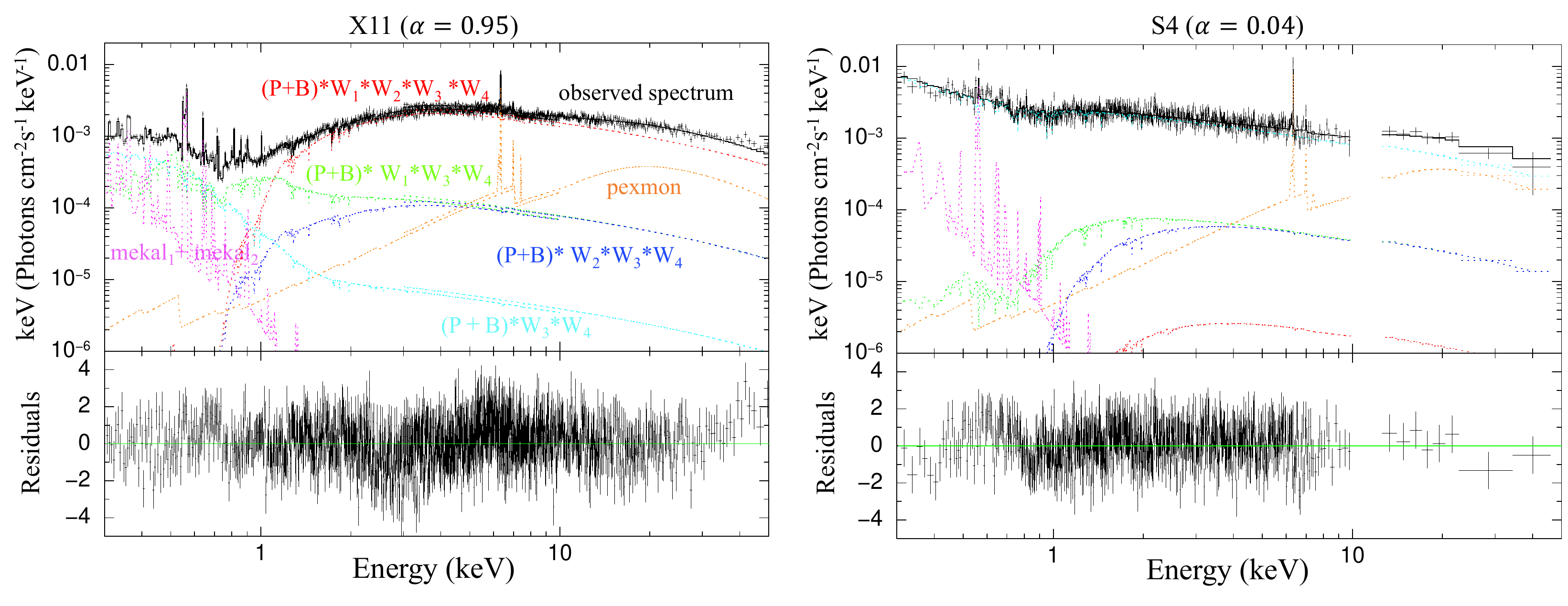}
	    \caption{Spectral fitting results of a representative simultaneous observation by \xmmnus (X11; left) and by \su XIS1-PIN (S4; right) within 0.3--50.0~keV. On the upper panel of both figures, black solid line is the best-fit model spectrum and the dotted-lines show individual continuum components. Those spectral components colored in light-blue, blue, green and red  correspond  to those with the same colors in the top-panel of Figure~\ref{fig:schematic}. Among all the four component going through the full absorbers ($W_3$ and $W_4$), the light-blue component is neither absorbed by the warm layer ($W_1$) nor the cold core ($W_2)$. Blue and green components are absorbed only by the cold core and the warm layer, respectively. Red one is absorbed by both absorbers. Note that the red component and the blue component are hardly distinguishable in shape, since $W_1$ is much optically thinner than $W_2$. Orange and magenta dotted-lines  describe the outer-disk reflection (\pexmon) and the thermal plasma component (two-temperature \mekal), respectively. The vertical axes show energy flux. 
	    The spectra are unfolded around the best-fit model using the XSPEC command {\tt eufspec}.} 
	    \label{fig:setpladd}
\end{figure*}

Therefore, we suppose that ``double-layer clumpy absorbers'' are more plausible, as in the lower illustration of Figure~\ref{fig:schematic}. In this case, it is naturally understood that the two absorbers have the same covering fraction. The only difference is in those X-rays indicated with the blue lines in the figure;
in the top panel, they are intervened only by the core of the clump ($W_2$), but they are necessarily intervened by the outer-layer of the clump ($W_1$) in the bottom panel.
However, since $W_1$ is much optically thinner than $W_2$, the additional $W_1$ absorption is not very significant (see Figure~\ref{fig:setpladd}). Thus, the double-layer absorbers in the bottom panel practically satisfy Equation~\ref{eq:vdpc}.

\subsection{Comparison with other objects}\label{sec:4.2}
 Previous studies suggest presence of the multi-layer/clumpy absorbers in other Seyfert galaxies besides \ngc. 
For example, NGC~3783 exhibits heavy eclipsing events, which can be naturally explained by two partially covering components with similar covering fractions \citep{Mehdipour17}.
In the case of NGC~1365, \citet{Maiolino10} proposed that partial absorbers are likely to be cometary-shape clouds, consisting of a high-density head and an elongate tail with lower density.
NGC~4051 is a bright AGN, in which  variability of the individual emission/absorption lines have
been studied in detail \citep{Mizumoto17}.
Its spectrum was explained by an invariable ionized absorber and variable double-layer clumpy absorbers, where the ionized absorber comprises two plasma components with different velocities and ionization states \citep{Mizumoto17}.
Thus, a picture of the ``invariable ionized absorber'' and the ``variable double-layer clumps'' seems to be common, while these absorbers' parameters are different depending on individual AGN circumstances.
Besides AGN, \citet{Hirsch19} found that the clumpy stellar winds of Cyg X-1 cause partial covering, where each clump consists of a cool and dense matter embedded in outer photoionized layer.

\subsection{Origin of the multiple absorbers}\label{sec:4.3}
We have seen that X-ray spectral variation of \ngc is naturally understood by introducing invariable ionized absorbers ($W_3$ and $W_4$) and variable double-layer clumpy absorbers ($W_1$ and $W_2$).
The former corresponds to the ``warm absorber'' \citep[WA; e.g.,][]{Kaastra00} and the latter corresponds to the ``obscurer'' \citep[e.g.,][]{Kaastra14} in previous studies.

It is being recognized that ``outflow'' is a common phenomenon in accreting systems such as AGNs and X-ray binaries. In fact, global radiation-magnetohydrodynamic simulations 
\citep[e.g., ][]{Takeuchi13,Kobayashi18} predict emergence of the clumpy outflow mostly due to the Rayleigh-Taylor instability. 
The disk wind outflow is hotter near the accretion disk, then cooled down as it moves farther away, transforming into the clumpy-outer wind at several hundred times the Schwarzschild radius. These clumpy-outer winds are likely origins of the double-layer obscurers.

The ``hot-inner and clumpy-outer wind'' picture successfully describes the characteristic broad Fe-K line shape and the time-lags commonly observed from narrow-line Seyfert galaxies such as 1H0707-495 \citep{Mizumoto19}. 
In the case of these high-mass accretion rate systems, the fast and thick hot-inner wind
would produce the blue-shifted Fe-K absorption lines \citep[e.g.,][]{2015MNRAS.446..663H} and/or scatter X-rays to produce P-Cygni like broad features \citep[e.g.,][]{2016MNRAS.461.3954H,Mizumoto19}.
On the other hand, such strong Fe-K features are not seen in the X-ray spectra of \ngc, presumably because the inner-wind may be fully ionized and/or optically thin.
Here, we instead suggest a possibility of the  thermal wind origin of the WA \citep[e.g.,][]{2001ApJ...561..684K,2019MNRAS.489.1152M}. Materials in the outer-torus or accretion flow are heated by X-rays, and evaporated to produce thermally driven outflows (Figure \ref{fig:otama}).
These ionized outflows with moderate velocities in the line-of-sight are presumably the WAs' origin to produce the X-ray and UV absorption lines.

The unified AGN model assumes the BLR clouds obscured by the dust torus (e.g., \citealp{Antonucci85}; \citealp{Urry95}). However, physical origin, geometry, and dynamics of the BLR clouds are still under discussion. Thus, relationship between the clumpy outer winds (obscures) of the VDPC model and the BLR clouds has not been fully elucidated.
On one hand, 
\citet{Miyakawa12} estimated the size and location of the partial absorbers and suggested that the partial covering clouds in the VDPC model are consistent with the BLR clouds.
Similarly, \citet{Matthews20} reproduced the BLR-like spectra assuming clumpy biconical disk winds illuminated by an AGN continuum.
\citet{Maiolino10} also claimed that the cometary clouds' parameters are consistent with those of the BLR clouds.
On the other hand, \citet{Kaastra14} argued that the obscurer in \ngc should be farther away from the BLR because it partially absorbed the broad emission lines.
According to the reverberation mapping observations of the STORM campaign (e.g., \citealp{Williams20}), the radius of the \ngc BLR is estimated to be several hundred times the Schwarzschild radius.
Simulation by \citet{Kobayashi18} predicted presence of the clumpy-outer winds farther than several hundred times the Schwarzschild radius, which corresponds to the position equal to or farther than the BLR.
Therefore, we suppose the clumpy absorbers in \ngc either correspond to the outer BLR clouds or are located beyond the BLR clouds.

\subsection{Variation of the spectral parameters over 16 years}\label{sec:4.4}
Figure~\ref{fig:setpladd} shows two representative energy spectra with similar continuum emissions but completely different partial covering fractions. The left shows the X11 data obtained by \xmmnus, and the right shows the S4 data obtained by \su XIS1-PIN. They have the same {\tt diskbb} normalization and low temperatures ($T_{\rm in} \approx$ 50--150~eV), which causes a minor spectral difference below $\sim$1~keV. 
Changes of the power-law normalization have little influence on the spectral shape.
Namely, spectral changes above $\sim$1~keV are essentially caused by only changes of the covering fraction.
The covering fractions of X11 and S4 are 0.95 and 0.04, respectively.
As seen in Figure~\ref{fig:setpladd}, the apparent change of the entire spectrum is mostly explained by only variation in the covering fraction.

In this study, the power-law photon index $\Gamma$ is invariable  between 2007 and 2016, except for X4. 
This  might sound inconsistent with the widely-accepted  ``softer-when-brighter'' trend  of Seyfert galaxies (e.g., \citealp{Sobolewska09}, \citealp{Soldi14}, \citealp{Liu21},\citealp{Laurenti22}). 
We point out that considering the partial absorption, larger covering fractions cause 
a reduction of the soft X-rays and  apparently flat   spectra  even when the intrinsic photon index is invariable. Thus, we suggest that the apparent
 softer-when-brighter trend is explained by a
 change of the partial covering fraction, at least to some extent.

The \diskbb flux $F_{\rm disk}$ is obtained by the relation $F_{\rm disk} = L_{\rm disk} \cos{i} /2\pi d^2$, where $L_{\rm disk} = 4\pi r_{\rm in}^2 \sigma T_{\rm in}^4$, $d$ is the distance to \ngc, 71.77~Mpc.
We also calculated \pl flux of the 0.3--50.0~keV band using the XSPEC {\tt flux} command.
These \diskbb and \pl fluxes are shown in Table~\ref{tab:var_par}, which may be combined to estimate the total unobscured continuum flux from the central region.
The \diskbb flux is significantly variable, but its spectrum is too soft to highly ionize those heavy elements that are ionized mainly by the hard power-law photons. As shown in Table~\ref{tab:var_par}, the \pl flux does not vary much, which justifies our early assumption that ionization parameters of the intervening matters are invariable.

\begin{figure}
        \includegraphics[width=1.0\columnwidth]{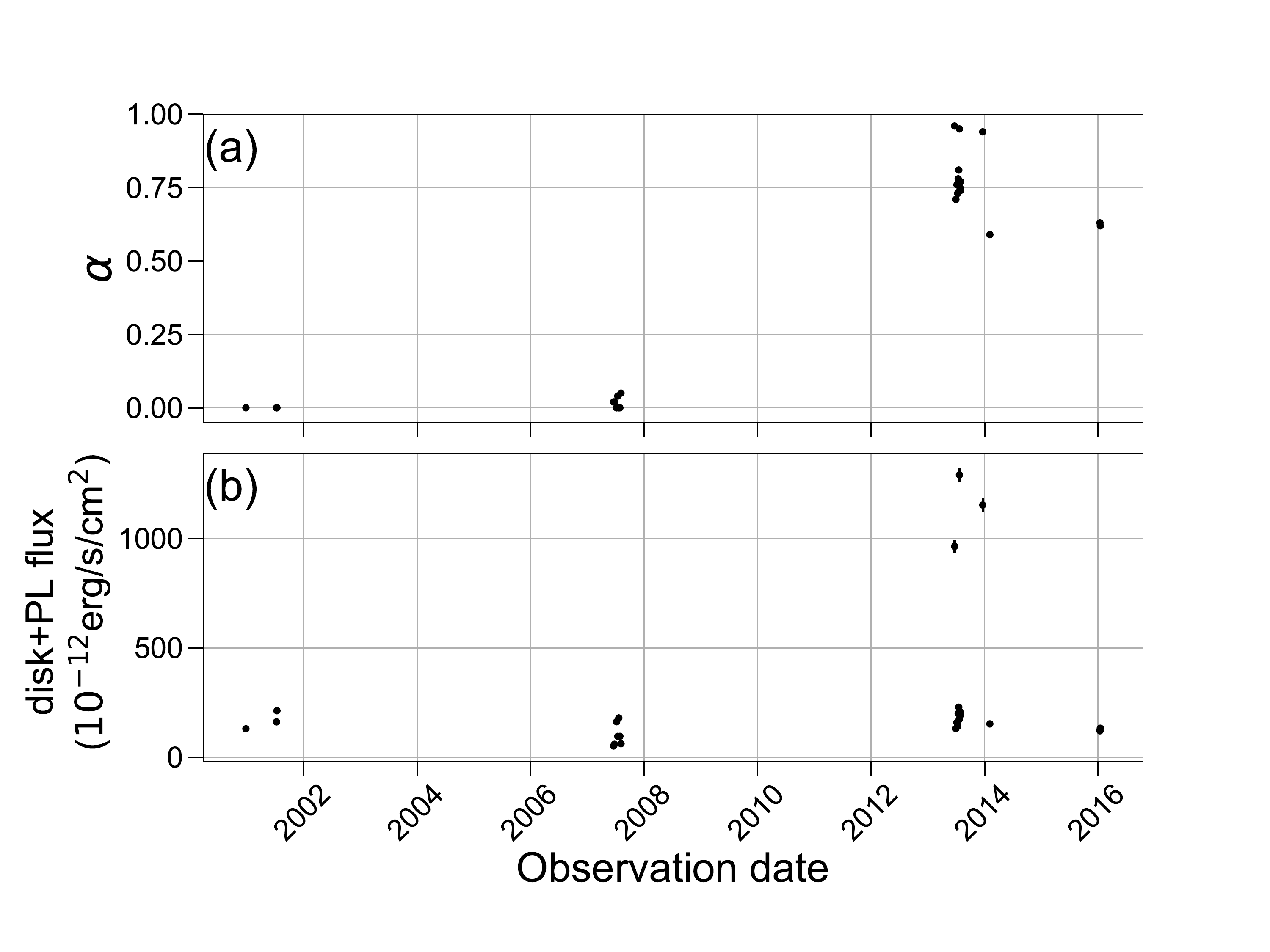}
	    \caption{Parameter variations during 2000--2016 shown in Table~\ref{tab:var_par}: (a) covering fraction, (b) sum of the unobscured \diskbb and \pl fluxes. 
	    }
	    \label{fig:par_var}
\end{figure}

\begin{figure}
        \includegraphics[width=1.0\columnwidth]{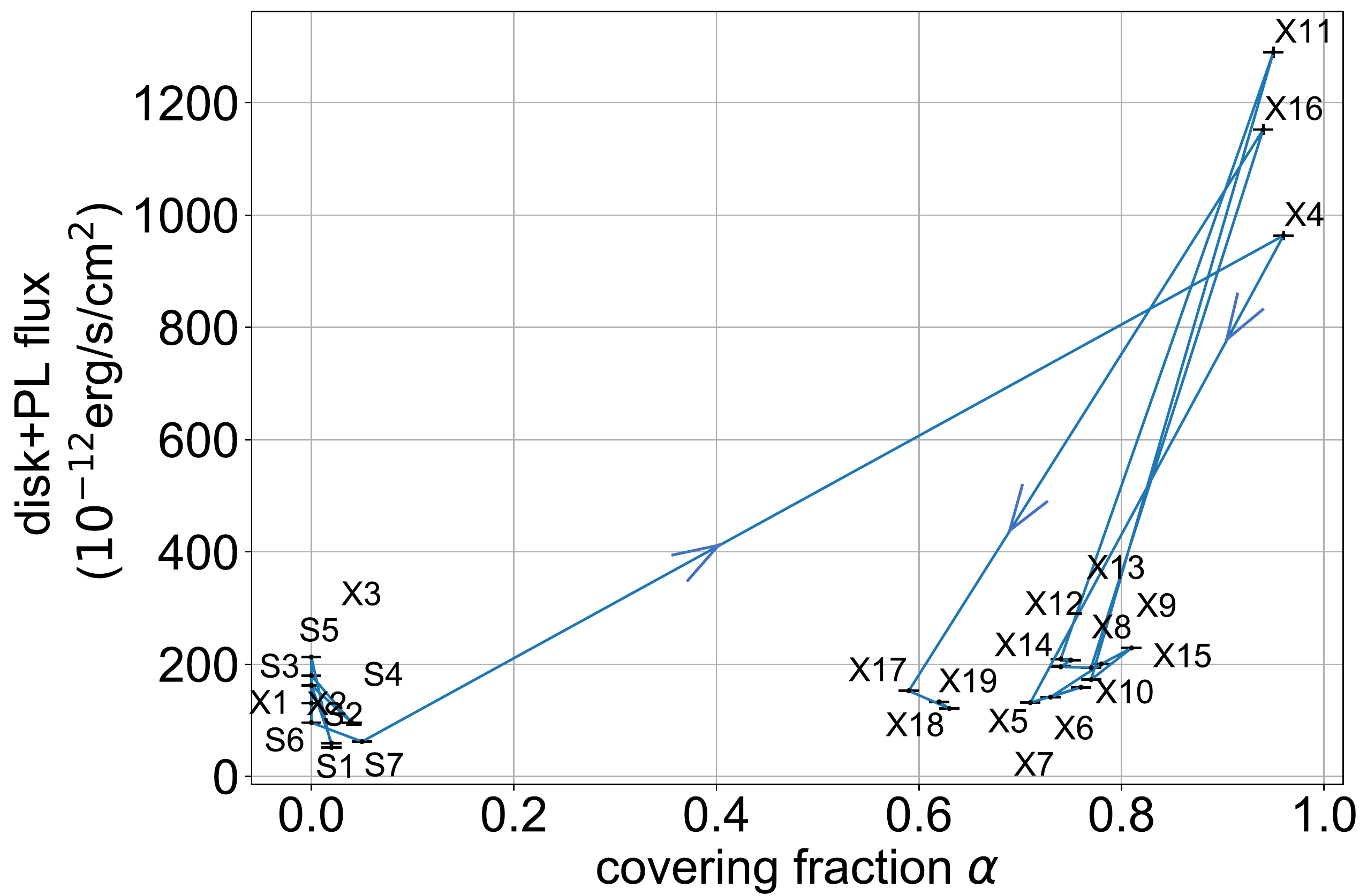}
	    \caption{Partial covering fraction versus sum of the unobscured \diskbb and \pl fluxes. The arrow indicates the time order.}
	    \label{fig:no_corr}
\end{figure}

Figure~\ref{fig:par_var} shows variations of the partial covering fraction and the total unobscured flux during 2000--2016. 
Relation between the two parameters is indicated in Figure~\ref{fig:no_corr}. 
We can see that the unobscured flux and the partial covering fraction are independent, which is reasonable since the former is determined by the inner accretion rate, and the latter represents the geometrical configuration of the clumpy absorbers.

\section{Conclusion}
We carried out a broad-band (0.3--78~keV) X-ray spectral analysis of \ngc using the
data obtained by \xmm, \nus, and \su in 26 observations from 2000 to 2016. We have found that the long-term spectral variations are explained by the ``variable double partial covering'' (VDPC) model, where variations of the partial covering fraction due to double-layer clumpy absorbers account for most of the spectral variations (Figure~\ref{fig:otama}). 
We propose a simple spectral variation model with only three variable components; the soft excess spectral component, the cut-off power-law normalization, and the partial covering fraction of the clumpy absorbers. 
Above $\sim$1~keV, the observed flux/spectral variation is explained by two essentially variable parameters; the partial covering fraction and the cut-off power-law normalization,
where the former dictates  change of the spectral shape.
In contrast,
the intrinsic photon index and all the other spectral parameters are not 
significantly  variable for over 16 years. 

\vspace{0.5cm}
\section*{Acknowledgements}
The authors gratefully thank the referee for critical comments to improve the paper.
This research has made use of data and software provided by the High Energy Astrophysics Science Archive Research Center (HEASARC), which is a service of the Astrophysics Science Division at NASA/GSFC and the High Energy Astrophysics Division of the Smithsonian Astrophysical Observatory.
This study was based on observations obtained with \xmm, ESA science missions with instruments and contributions directly funded by ESA Member States and NASA.
This research has made use of data obtained with the \nus mission, a project led by the California Institute of Technology (Caltech), managed by the Jet Propulsion Laboratory (JPL) and funded by NASA.
We utilized the public \su data obtained by the Data ARchives and Transmission System (DARTS) provided by the Institute of Space and Astronautical Science (ISAS) at the Japan Aerospace Exploration Agency (JAXA).
This research was supported by JSPS Grant-in-Aid for JSPS Research Fellow Grant Number JP20J20809 (T.M.), JSPS KAKENHI Grant Number JP16K05309 (K.E.) and JP21K13958 (M.M). 
M.M. acknowledges support from the Hakubi project at Kyoto University. 
The authors acknowledge Ms.\ Juriko Ebisawa for proofreading and the artwork for Figure \ref{fig:otama}.

\section*{Data Availability}
Observational data used in this paper are publicly available, and any additional data will be available on request.




\bibliographystyle{mnras}
\bibliography{5548zotero}

\section*{Appendix A: Energy spectral fitting}
\setcounter{section}{1} 
\setcounter{figure}{0} 
\renewcommand{\thefigure}{\Alph{section}.\arabic{figure}}

As discussed in Section~\ref{sec:3.3}, we performed energy spectral fitting with all the available archival data. Figure~\ref{fig:allxmm_fit} and Figure~\ref{fig:allsu_fit} show the results of the model fitting. The best-fit parameters are shown in Table~\ref{tab:var_par}.

\begin{figure*}
\begin{center}
        \includegraphics[width=1.8\columnwidth]{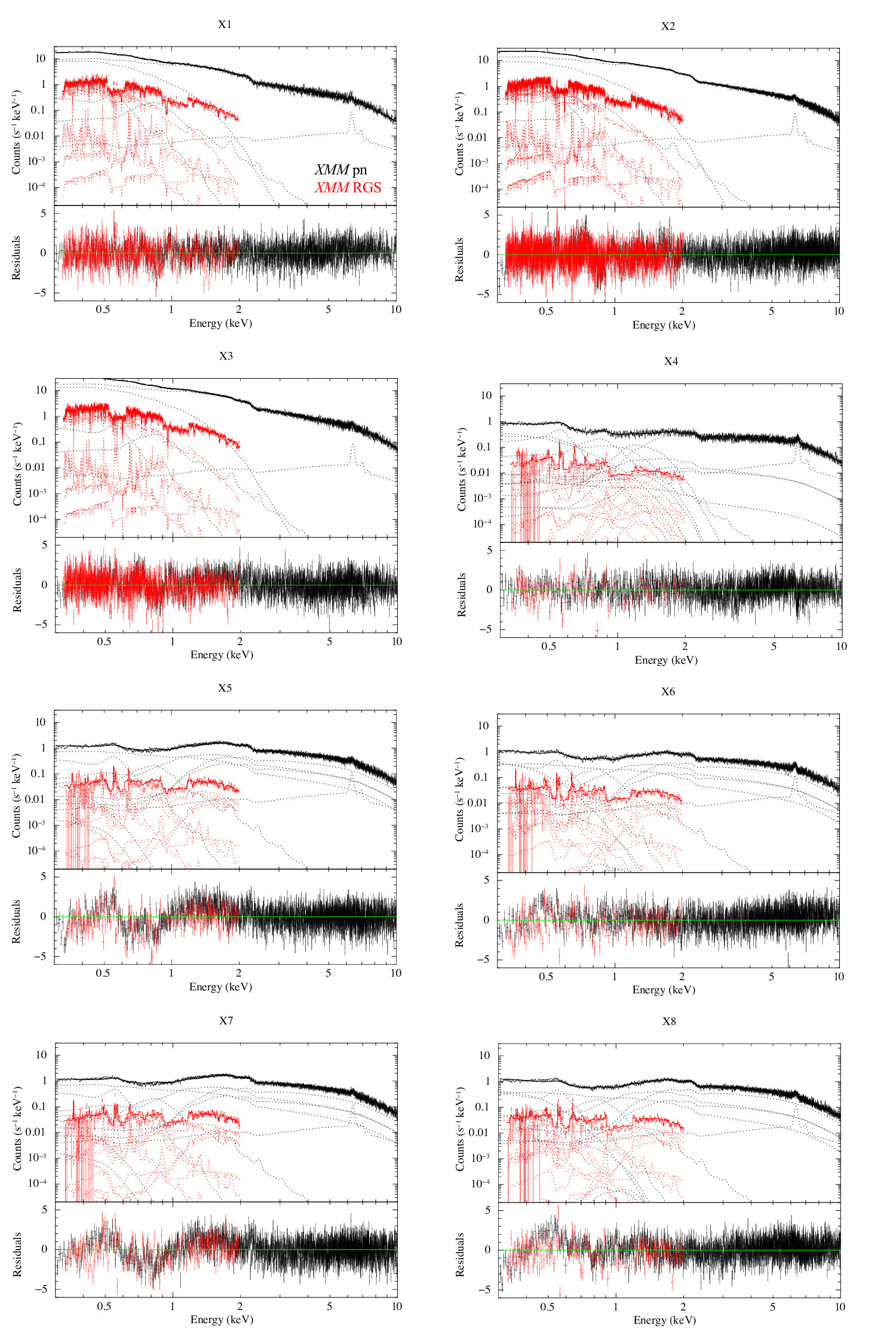}
	    \caption{Spectral fitting results of all the spectra taken by \xmm. Black and red lines show the pn and RGS spectra, respectively. The vertical axes in both the panels are the same as those in Figure~\ref{fig:4obs}}
	    \label{fig:allxmm_fit}
\end{center}
\end{figure*}

\addtocounter{figure}{-1}
\begin{figure*}
\begin{center}
  \includegraphics[width=1.8\columnwidth]{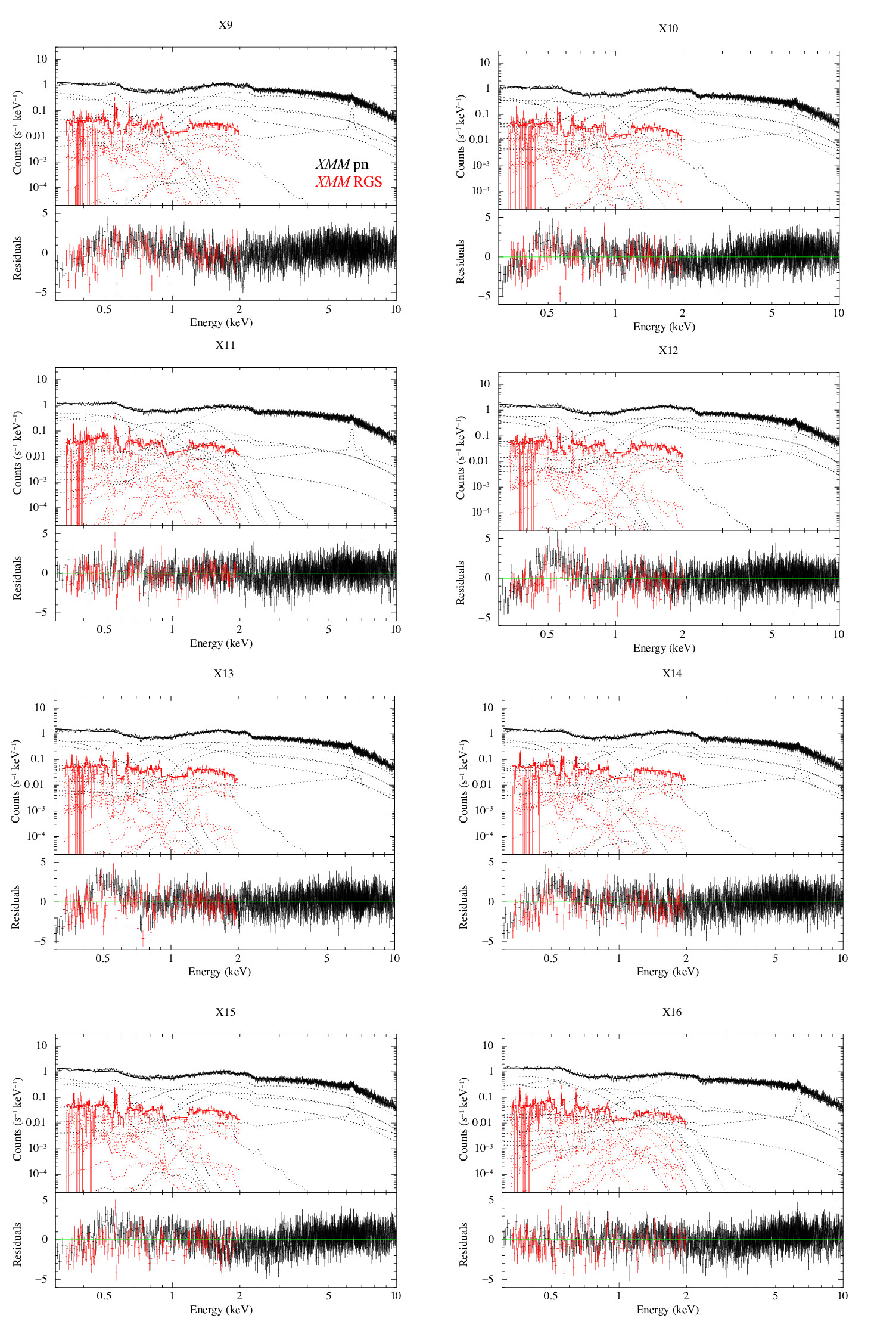}
  \caption[]{Cont.}
\end{center}
\end{figure*}

\addtocounter{figure}{-1}
\begin{figure*}
\begin{center}
  \includegraphics[width=1.8\columnwidth]{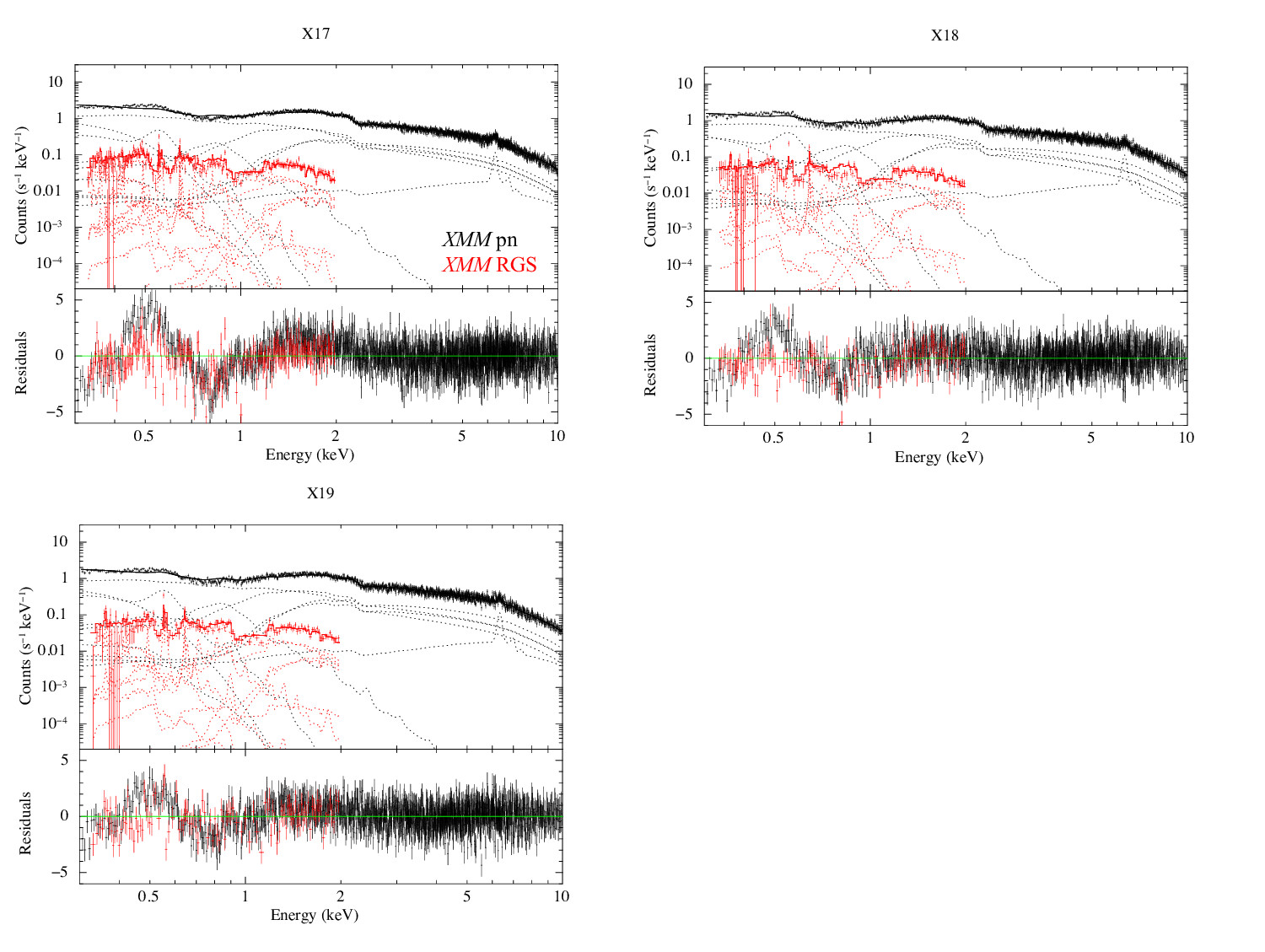}
  \caption[]{Cont.}
\end{center}
\end{figure*}

\begin{figure*}
\begin{center}
  \includegraphics[width=1.8\columnwidth]{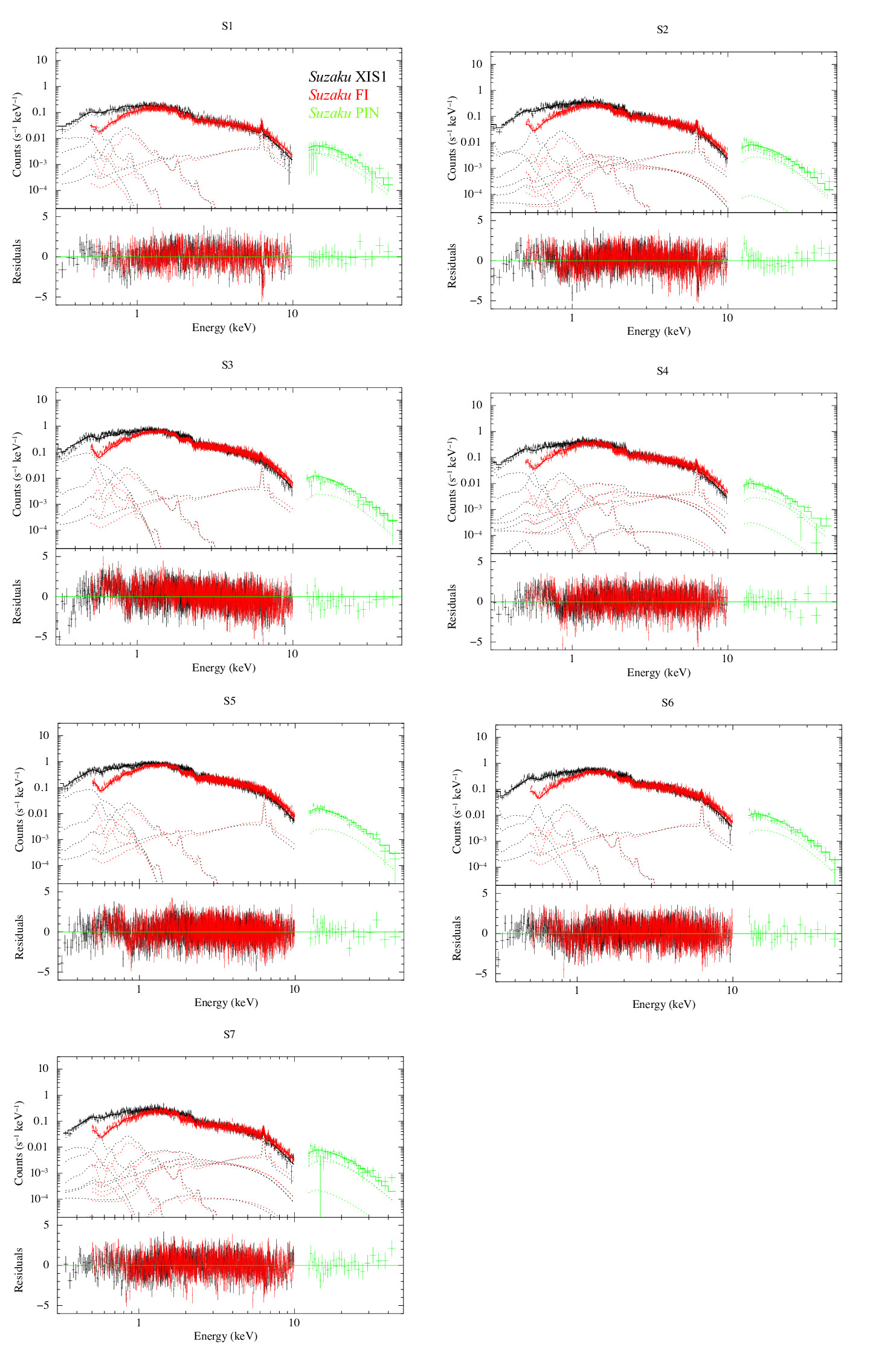}
  \caption{Spectral fitting results of all the spectra taken by \su. Black, red, and green lines show the XIS1, FI (XIS0\&3), and PIN spectra, respectively. The vertical axes in both the panels are the same as those in Figure~\ref{fig:4obs}}
  \label{fig:allsu_fit}
\end{center}
\end{figure*}


\bsp	
\label{lastpage}
\end{document}